\documentclass[aps,prx,twocolumn,superscriptaddress]{revtex4-2}

\usepackage{graphicx}
\usepackage{amsmath}
\usepackage{amssymb}
\usepackage{float}
\usepackage[colorlinks,urlcolor=blue]{hyperref}
\usepackage[capitalise]{cleveref}
\usepackage{nicefrac}
\usepackage{multirow}
\usepackage[table]{xcolor}

\usepackage{bigints}
\usepackage{enumitem}

\definecolor{lightmagenta}{RGB}{255,100,200}
\def\doubleunderline#1{\underline{\underline{#1}}}

\newcommand{\SM}[1]{{\color{black}{#1}}}

\begin{document}
\title{Experimentally Motivated Order of Length Scales Affect Shot Noise}
\author{Sourav Manna}
\email{sourav.manna@weizmann.ac.il}
\affiliation{Department of Condensed Matter Physics, Weizmann Institute of Science,
Rehovot 7610001, Israel}
\affiliation{Raymond and Beverly Sackler School of Physics and Astronomy, Tel-Aviv
University, Tel Aviv 6997801, Israel}	
\author{Ankur Das}
\email{ankur@labs.iisertirupati.ac.in}
\affiliation{Department of Physics, Indian Institute of Science Education and Research (IISER) Tirupati, Tirupati 517619, India}
\affiliation{Department of Condensed Matter Physics, Weizmann Institute of Science,
Rehovot 7610001, Israel}

\begin{abstract}
Shot noise at a conductance plateau in a quantum point contact (QPC) can be explained by
considering equilibrations at the quantum Hall edges. The indication from recent experiments is that the charge equilibration length is much shorter than the thermal equilibration length.
We discuss how this discovery gives rise to different 
thermal equilibration regimes in the presence of full
charge equilibration. 
In this work, we classify these distinct
regimes via dc current-current correlations \emph{(electrical
shot noise)} at definite experimentally found (or possible) QPC conductance plateaus for the edges of integer, particle-like,
and hole-like filling fractions in a 
two dimensional electron gas.
Our analyses show that distinct universal features arise among the different
thermal equilibration regimes for the edges of particle-like and
hole-like states.
\end{abstract}

\maketitle

\section{Introduction}
\label{sec:intro}
Quantum Hall effect is one of the oldest known phenomena that requires
understanding of band topology and quantum mechanics \cite{PhysRevLett.48.1559, PhysRevLett.50.1395, RevModPhys.82.3045, RevModPhys.89.025005, RevModPhys.95.011002}. 
However, the bulk-boundary correspondence \cite{PhysRevB.43.11025} does not guarantee a comprehensive understanding of the edges for a given bulk
topological order.
While the bulk is gapped, the edges are gapless which 
participate in both electrical and thermal transport carrying both
charge and energy respectively. The edge modes can be either co-propagating or counter-propagating and can exhibit a distinct
steady state.
Thus, we can separate our question of the quantum Hall effect into two different pieces: (1) the topological order
of the bulk and (2) the details (for example the steady state) of the edge.  There is more than one known
mechanism that can make the picture of the edge ``fuzzy" keeping the bulk-boundary correspondence intact.

One of the phenomena that plagued the quantum Hall edge theory is the edge reconstruction.
This is an ``umbrella-term" that includes the appearance of new counter-propagating modes \cite{Wang2013,PhysRevB.103.L121302, PhysRevLett.129.146801}, reordering of the edge modes \cite{PhysRevLett.119.186804}, emergence of an effective
narrow strip of spin rotation \cite{PhysRevLett.128.186801}, and more.
This was first discussed for the two dimensional electron gas (2DEG)
by Chamon and Wen \cite{PhysRevB.49.8227} almost three decades ago.
They explained how the sharpness of the confining potential can
cause the appearance of new counter-propagating edge modes.
Following this,
there have been numerous efforts to understand the edge reconstruction for integer 
and fractional quantum Hall systems \cite{Khanna_2022}. More recently, we have learned that even integer quantum
Hall edges can have a fractional edge reconstruction \cite{PhysRevB.103.L121302}. The general answer to the question of when and 
how edge reconstruction takes place is a non-universal one. This makes the study of
edge reconstruction a difficult and detailed topic of interest for both the experimentalists and theorists.

Typically in
an experiment, a dc voltage is used as the source which causes voltage drops
near different contacts. This
gives rise to the Joule heating
which produces heat and the
heating effects can play a crucial role in the edge transport.
Heat, which propagates along the edge modes, induces particle-hole pairs and
thermally activates the tunneling of particles and holes among the edge modes. This gives rise to a remarkable phenomenon known as the edge equilibration \cite{Protopopov2017, PhysRevLett.123.137701, PhysRevB.99.161302, PhysRevLett.125.157702, SM5by2, PhysRevB.98.115408, PhysRevB.101.075308, PhysRevB.104.115416}. Such an incoherent phenomenon leads the steady state of edges to a hydrodynamic 
regime and the edge modes are equilibrated 
to a hydrodynamic mode
of a given chirality. Such a mode is characterized by its electrical and thermal conductances. Tunneling of particles and holes allows the exchange of both charge and heat. This leads to the electro-chemical potential and thermal equilibrations, respectively \cite{PhysRevB.101.075308,Protopopov2017,PhysRevB.98.115408}.
The particle-hole pairs can split stochastically into two modes and move into opposite directions in the case of the edges comprising of counter-propagating modes. This mechanism gives rise to the electrical shot noise (current-current correlations), 
which provides a fully electrical approach to capture the effect of heat propagation along the edges.

\begin{figure}[t]
\includegraphics[width=\columnwidth]{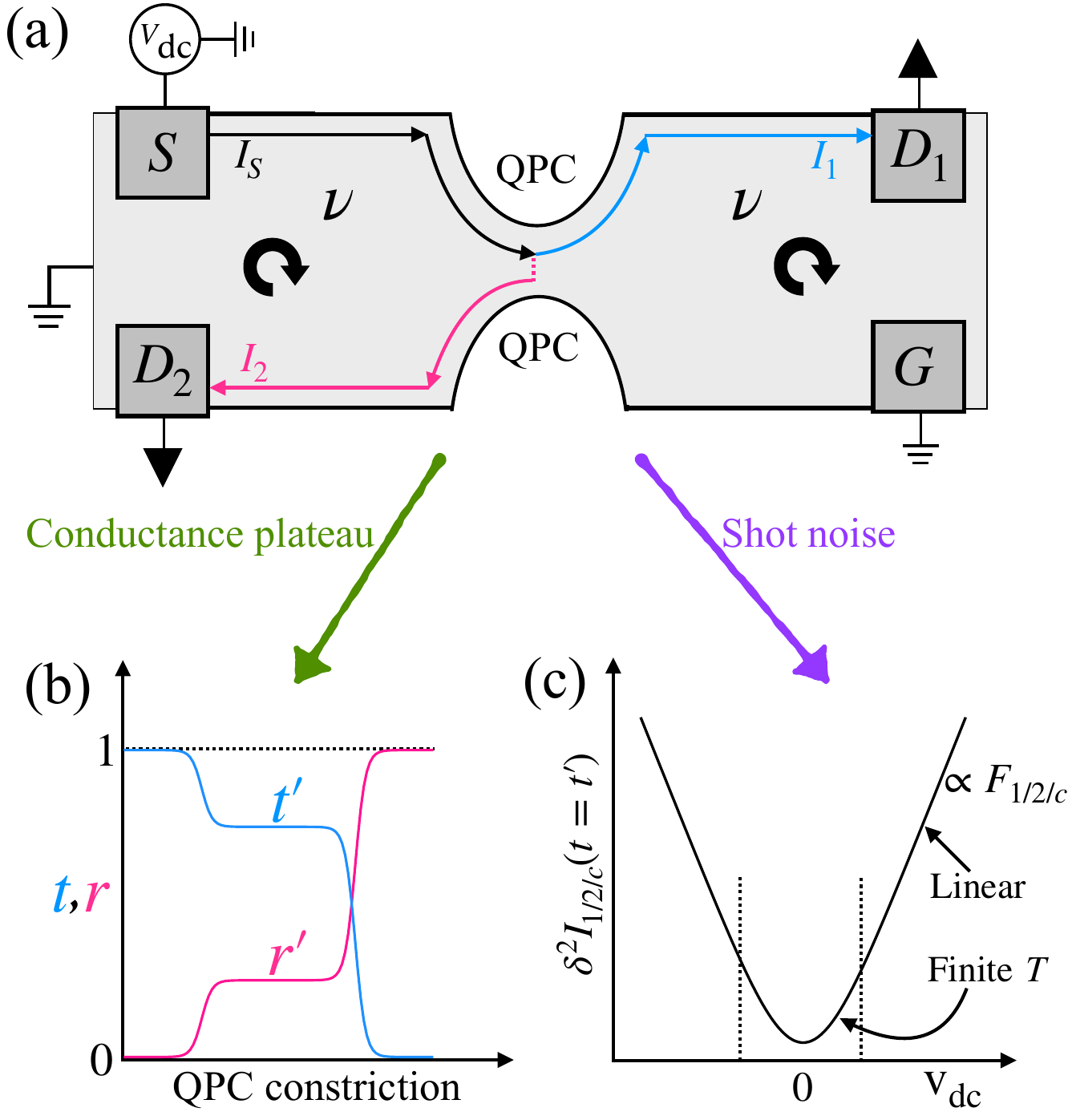}
\caption{A schematic picture of the device and measurements, which we use
throughout the paper. (a) We show a Hall bar with filling $\nu$ in a QPC geometry.
We have four contacts as a source $S$, a ground $G$, and two drains $D_1, D_2$. A dc
current $I_S$ is injected by $S$, which is biased by a dc voltage $V_{\text{dc}}$, and
the measurements are performed at $D_1, D_2$. We describe the chirality of charge
propagation by circular arrows. (b) Transport measurements are carried out by measuring
currents $I_1, I_2$ at $D_1, D_2$ respectively. The transmission and reflection
coefficients are $t=I_1/I_S$ and $r=I_2/I_S$ respectively, where $t+r=1$. A QPC transmission
plateau can be observed at a transmission $t=t'$ while measuring $t$ (correspondingly $r$) as a function of the QPC constriction (from a fully open to a fully pinched off
QPC). (c) At a transmission plateau ($t=t'$), the current reaching $D_1, D_2$ can
be noisy (leading to $\delta^2_{I_{1/2/c}}(t=t') \neq 0$)
and there can be
different mechanisms to it. The functional
dependence of $\delta^2_{I_{1/2/c}}(t=t')$ is schematically shown while changing $V_{\text{dc}}$
and the slope of the linear region is proportional to the corresponding Fano factor $F_{1/2/c}$. The curvature in $\delta^2_{I_{1/2/c}}(t=t')$ shows up due to the temperature dependence.} \label{Cartoon}
\end{figure}

In electrical current, there can be fluctuations due to the stochastic
injection of particles or holes. Now, the topological properties of a
quantum Hall edge allows only certain types of quasi-particle tunnelings to take
place, which in turn allows the measurement of quasi-particle charge. Quantum point contact (QPC), described below (c.f.\ \cref{Cartoon}), is one of the ways
to generate this stochastic tunneling \cite{PhysRevB.45.1742,MartinNoiseReview,Mahalu1997,PhysRevLett.79.2526,Mahalu1998,Saminadayar1998,Dima2023,PhysRevLett.102.106403} . However, this is not the
only mechanism that can generate shot noise. For example, the decay of
neutralons (quanta of the 
neutral modes) can be one of those mechanisms \cite{Park2021,Biswas2022,PhysRevB.103.L121302,Sabo2017}. We will discuss another mechanism due to the equilibration later.

Now, the natural question will be how to poke the
properties of the edge modes beyond the topological charge and thermal conductances. One of the
established techniques is to push the opposite edges of a quantum Hall system close to each
other so that they start to tunnel. This is done by an electrostatic constriction, 
namely a QPC. In the context of the quantum Hall effect, a QPC plays a pivotal role in 
understanding the microscopic structure of the edge. There are 
at least two types of measurements one can do in a
QPC device : (1) transport and (2) shot noise. As the QPC constriction gets narrower,
less current will transmit across the QPC. Now one question can be asked if we find that the 
transmission remains constant (leading to a transmission plateau) as we make the constriction narrower \cite{Bid2010,Bhattacharyya2019}. In that case, there can be at
least two possible scenarios: (1) there are more than one coherently propagating edge modes,
and a few of them completely backscatter, while the others fully transmit across the QPC and (2) at the QPC constriction, a very small region of a quantum
Hall liquid of a different filling fraction is stabilized such that when these modes 
electrically equilibrate (incoherent regime), it leads to a transmission plateau. We can also
ask if there are any shot noise signatures of these scenarios other than the transport measurements.

In recent times, there have been
immense development in QPC and sample standardizations and we have learned that we must treat the 
equilibration scenario more carefully. What we have learned is that there are \emph{two different} internal
length scales involved here, namely the electro-chemical potential or charge equilibration length 
($l^\text{ch}_\text{eq}$) and the thermal equilibration length ($l^\text{th}_\text{eq}$).
Latest
experiments have demonstrated that they are not of the same order of magnitude \cite{PhysRevLett.126.216803, Melcer2022, Kumar2022, Srivastav2022}. There are two
geometric length scales in a QPC geometry (c.f.\ \cref{Cartoon}): (1) length of the arms ($L_\text{A}$) and (2) size of the QPC 
($L_\text{Q}$). We know that $L_\text{A}\gg L_\text{Q}$ and from the experiments \cite{PhysRevLett.126.216803, Melcer2022, Kumar2022, Srivastav2022} we have
$l^\text{ch}_\text{eq} \ll l^\text{th}_\text{eq}$.
To have a transimission plateau we must have $l^\text{ch}_\text{eq} \ll 
(L_\text{A}, L_\text{Q})$ leading to the full charge equilibration in every segment of the QPC geometry.
However, the degree of thermal equilibration can be in a distinct regime in different segments.
Thus, the interplay among the order of magnitudes of $l^\text{th}_\text{eq},\ L_\text{A},\ L_\text{Q}$ leads to distinct
inequalities giving rise to a zoo of scenarios.
We study the effect of these inequalities in the electrical
shot noise (defined below).\SM{
We point out that the present work goes significantly beyond our earlier works \cite{SM5by2,SMLong,SMshort} as follows.
Ref.\ \onlinecite{SM5by2} deals with a
device based on interfacing different quantum Hall
filling---thereby the electrical conductance plateaus
are defined by the gated regions, whereas in our
present work we study a QPC
geometry where an electrical conductance plateau 
originates from the edge structure of the corresponding
bulk filling. We considered only full and partial
thermal equilibrations in Ref.\ \onlinecite{SM5by2}, eliminating any effect of edge reconstruction, whereas in our 
present work we considered no, hybrid ($L_\text{Q} \ll l^\text{th}_\text{eq} \ll L_\text{A}$), and full thermal
equilibrations and thereby the effect of edge reconstruction is present. 
Notably, the hybrid thermal equilibration scenario is not possible in Ref.\ \onlinecite{SM5by2} since we considered an interface-based device there.
We considered the $5/2$ bulk filling in Ref.\ \onlinecite{SM5by2}, which is a particle-hole symmetric state and here, we have studied the states having no particle-hole symmetry.
In Refs.\ \onlinecite{SMLong,SMshort}, we considered only $2/3$ quantum Hall
bulk filling, whereas in our present work we go beyond
that and study other Abelian bulk filling factors.
}

We refer to \cref{Cartoon} for a schematic picture of the QPC geometry and the corresponding transport
and shot noise
measurements. We have $I_1$ and $I_2$ as the currents which enter the drains $D_1$ and
$D_2$, respectively. We define the dc current-current auto-correlations 
as $\delta^2 I_1$
in $D_1$, $\delta^2 I_2$ in $D_2$, and the cross-correlation as $\delta^2 I_c$
\cite{averagedef} where
\begin{equation}
\begin{split}
&\delta^2 I_1=\langle (I_1 - \langle I_1 \rangle)^2 \rangle,\\&\delta^2 I_2=
\langle (I_2 - \langle I_2 \rangle)^2 \rangle,\\&\delta^2 I_c=\left\langle
\left(I_1 - \langle I_1 \rangle\right)
\left(I_2 - \langle I_2 \rangle\right) \right\rangle.
\end{split}
\end{equation}
The shot noise Fano factors are defined as 
\begin{equation}\label{Fano}
    \begin{split}
        F_j = \frac{\delta^2 I_j}{2e I_S t(1-t)},\ j \in \{1,2,c\}
    \end{split}
\end{equation}
where $I_S$ is the source current and $t = I_1/I_S$ is the QPC transmission.

In the following sections, we describe the keynote points of our work
(\cref{sec:achieve}), the assumptions and techniques we use (\cref{sec:assum}),
and our insight into the results (\cref{sec:results}). We draw a summary and the future prospect
of our work in \cref{sec:summary}.

\section{Keynote points}
\label{sec:achieve}
Being motivated by the experiments \cite{Bhattacharyya2019,Manfra2022} and previously
studied cases \cite{SM5by2,SMshort,SMLong}, we analyze the effects of different orders
of length scales on the shot noise signatures and compare the corresponding different Fano factors in full
completeness. Here, we not only do a qualitative analysis 
but also calculate the quantitative values of the Fano factors for all
the cases considering the inequalities among $l^\text{ch}_\text{eq}$, $l^\text{th}_\text{eq}, L_\text{A}, L_\text{Q}$. As we do not have any control over the internal length scales,
we consider different reasonable scenarios.
\SM{Below we discuss the keynote points of our
results 
relating one another and explicitly stressing their importance.}

\begin{table*}[t!]
  		\begin{tabular}{| c | c || c || c || c |}
			\hline
            $\nu$ & $\nu_i$  & No equilibration $(F_1=F_2, F_c)$ & Hybrid equilibration $(F_1=F_2, F_c)$ & Full equilibration $(F_1=F_2, F_c)$ \\ \hline \hline
			\multirow{2}{*}{$1$}     & $1/3$ & $\approx (0.36, -0.36)$ \cellcolor{lightmagenta} & $\approx (0.36, -0.36)$ \cellcolor{lightmagenta} & $\approx ( 0.55\sqrt{l_{\text{eq}}^\text{th}/L_\text{Q}}, -0.55\sqrt{l_{\text{eq}}^\text{th}/L_\text{Q}})$ \\ \cline{2-5}
			& $2/3$ & $\approx (0.33, -0.33)$ \cellcolor{lightmagenta} & $\approx (0.33, -0.33)$ \cellcolor{lightmagenta}     & $\approx ( 0.55\sqrt{l_{\text{eq}}^\text{th}/L_\text{Q}}, -0.55\sqrt{l_{\text{eq}}^\text{th}/L_\text{Q}})$ \\ \cline{2-5}
            \hline\hline
            \multirow{1}{*}{$2/5$}     & $1/3$ & $\approx (0.1, -0.1)$\cellcolor{lightmagenta} & $\approx (0.1, -0.1)$\cellcolor{lightmagenta}& $\approx (0.0, -0.0)$ \\ \cline{2-5}
            \hline\hline
            \multirow{2}{*}{$3/7$}     & $1/3$ & $\approx (0.086, -0.086)$\cellcolor{lightmagenta} & $\approx (0.086, -0.086)$\cellcolor{lightmagenta}& $\approx (0.0, -0.0)$ \\ \cline{2-5}
            & $2/5$ & $\approx (0.066, -0.066)$\cellcolor{lightmagenta} & $\approx (0.066, -0.066)$\cellcolor{lightmagenta}& $\approx (0.0, -0.0)$ \\ \cline{2-5}
            \hline\hline
            \multirow{2}{*}{$2/3$}     & $1/3$ & $\approx (0.35, -0.22)$ & $\approx ( 0.09+0.3\sqrt{L_\text{A}/l^\text{th}_{\text{eq}}}, 0.09-0.3\sqrt{L_\text{A}/l^\text{th}_{\text{eq}}})$\cellcolor{cyan} & $\approx ( 0.09+0.3\sqrt{L_\text{A}/l^\text{th}_{\text{eq}}}, 0.09-0.3\sqrt{L_\text{A}/l^\text{th}_{\text{eq}}})$\cellcolor{cyan} \\ \cline{2-5}
             & $[1/3]$ & $\approx (0.28, -0.1)$ & $\approx ( 0.09+0.3\sqrt{L_\text{A}/l^\text{th}_{\text{eq}}}, 0.09-0.3\sqrt{L_\text{A}/l^\text{th}_{\text{eq}}})$\cellcolor{cyan}& $\approx ( 0.09+0.3\sqrt{L_\text{A}/l^\text{th}_{\text{eq}}}, 0.09-0.3\sqrt{L_\text{A}/l^\text{th}_{\text{eq}}})$\cellcolor{cyan} \\ \cline{2-5}
            \hline\hline
            \multirow{2}{*}{$3/5$}     & $1/3$ & $\approx (0.37, -0.097)$ & $\approx (0.53, -0.31)$\cellcolor{cyan}& $\approx (0.53, -0.31)$\cellcolor{cyan} \\ \cline{2-5}
            & $2/5$ & $\approx (0.38, -0.09)$ & $\approx (0.52, -0.3)$\cellcolor{cyan}& $\approx (0.52, -0.3)$\cellcolor{cyan} \\ \cline{2-5}
            \hline\hline
            \multirow{3}{*}{$4/7$}     & $1/3$ & $\approx (0.39, -0.01)$ & $\approx (0.39, -0.19)$\cellcolor{cyan}& $\approx (0.39, -0.19)$\cellcolor{cyan} \\ \cline{2-5}
            & $2/5$ & $\approx (0.43, -0.009)$ & $\approx ( 0.39, -0.18)$\cellcolor{cyan}& $\approx (0.39, -0.18)$\cellcolor{cyan} \\ \cline{2-5}
            & $3/7$ & $\approx (0.45, -0.008)$ & $\approx (0.39, -0.17)$\cellcolor{cyan}& $\approx (0.39, -0.17)$\cellcolor{cyan} \\ \cline{2-5}
            \hline
		\end{tabular}
		\caption{Summary of our shot noise results : 
$F_1, F_2$ are the auto-correlation Fano factors for the drains $D_1, D_2$, respectively,
while $F_c$ is the cross-correlation Fano factor for 
different bulk filling
factor $\nu$ and QPC filling factor $\nu_i (< \nu)$ in
distinct thermal equilibration  
regimes namely ``No", ``Hybrid", and ``Full".
We assume full charge equilibration 
and different degree of thermal equilibration in each of the
segments $L_{\text{A}}, L_{\text{Q}}$ (c.f.\ \cref{GeometryAndEff}).
We always have $F_1=F_2$, but $F_c$ is not always equal in magnitude to $F_1=F_2$. For the $\nu=$ particle-like states, ``No" and ``Hybrid" thermal equilibrations have the same values for $F_1=F_2, F_c$ (shaded pink) for a given 
$\{\nu, \nu_i\}$. Similarly, for the $\nu=$ hole-like states we find that $F_1=F_2, F_c$ acquire the same values for ``Hybrid" and ``Full" thermal equilibrations (shaded blue) for a given 
$\{\nu, \nu_i\}$. Here, $l^\text{th}_\text{eq}$ 
denotes the thermal equilibration length and $\approx 0$ indicating an exponential suppression as
a function of the geometric lengths. We mention that $\{\nu, \nu_i\} = \{2/3, [1/3]\}$ denotes the $1/3$ edge reconstruction in $\nu=2/3$ \cite{Wang2013} and the results 
for $\nu=2/3$ also appear in \cite{SMshort,SMLong}.
\SM{For a given bulk filling fraction, either $3/5$ or $4/7$, we note the following. Corresponding to a given QPC filling, while $F_1=F_2$ are numerically close for different thermal equilibration regimes, however, $F_c$ differs significantly
by order of magnitude. Hence, this can facilitate the 
experimental detection. Notably, $F_c$ is 
routinely measured in different experiments \cite{FeveColl,Lee2023}.}}
\label{tab:CCC_Table}
\end{table*}

\SM{
Depending on whether the quantum Hall bulk filling is particle-like or hole-like, the edge contains 
either co-propagating or counter-propagating modes
respectively, arising from the topology.
Naturally, question arises whether such difference
in the edge structure leaves its impressions in shot
noise in different equilibration regimes.
Indeed, 
in these results (c.f.\ \cref{tab:CCC_Table}), we find interesting
structures in the Fano factors for different regimes. We find that the $\nu=$
particle-like and $\nu=$ hole-like states (c.f.\ \cref{FFEdges}) behave \emph{differently}
in their shot noise signatures. 

Immediate consequences of different regimes of thermal
equilibrations are manifested in the shot noise for 
distinct bulk fillings---particle-like or hole-like.
Namely, the Fano factor
values for the $\nu=$ particle-like states are the same for the ``No" and 
``Hybrid" thermal equilibrations (arms are 
thermally equilibrated while edges around the QPC 
are not thermally equilibrated) 
for a given 
$\{\nu, \nu_i\}$, where
$\nu_i(<\nu)$ is the QPC filling.
This is because all the modes are
co-propagating, and equilibrating those in the arm (``outer" segment) does not make a difference (c.f.\ \cref{ParticleAndHole}(a)).
However, the situation changes for the $\nu=$ hole-like 
states where the Fano factors for the ``Hybrid" and ``Full" thermal equilibrations are equal
for a given 
$\{\nu, \nu_i\}$. 
This is because the counter-propagating 
modes 
in the arm (``outer" segment)
make sure that all the heat from a hot spot reaches the noise spots for both the 
``Hybrid" and ``Full" thermal equilibration regimes (c.f.\ \cref{ParticleAndHole}(b)). 

Now, the degree of thermal equilibration also leaves a profound
impact on the heat transport of a given segment, which
in turn shows up on the geometric length dependence of
the shot noise.
The Fano factors can be exponentially
suppressed (as a function of the geometric
lengths) or constant or length dependent
depending on the nature of heat transport
(ballistic or
antiballistic or diffusive, respectively) in different segments of the QPC geometry. 
Moreover, due to the antiballistic heat transport 
we also find that for all the $\nu=$
hole-like states, the auto- and the cross-correlation Fano factors are \emph{not equal}.
Though it is not expected from the conventional picture of the shot noise at a QPC
geometry but can be understood in the equilibration picture.

Thus, this work provides a comprehensive study,
as summarized in \cref{tab:CCC_Table}, of the shot noise Fano factors
where (1) we take the order of 
internal lengths $l^\text{ch}_\text{eq}$, $l^\text{th}_\text{eq}$
seriously including their comparison with the geometric lengths $L_\text{A}, L_\text{Q}$ and (2) we find qualitative
inequalities for the different Fano factors
(besides quantitative values)
resolving distinct scenarios.}

\section{Assumptions and techniques}
\label{sec:assum}
\begin{figure}
	\includegraphics[width=\columnwidth]{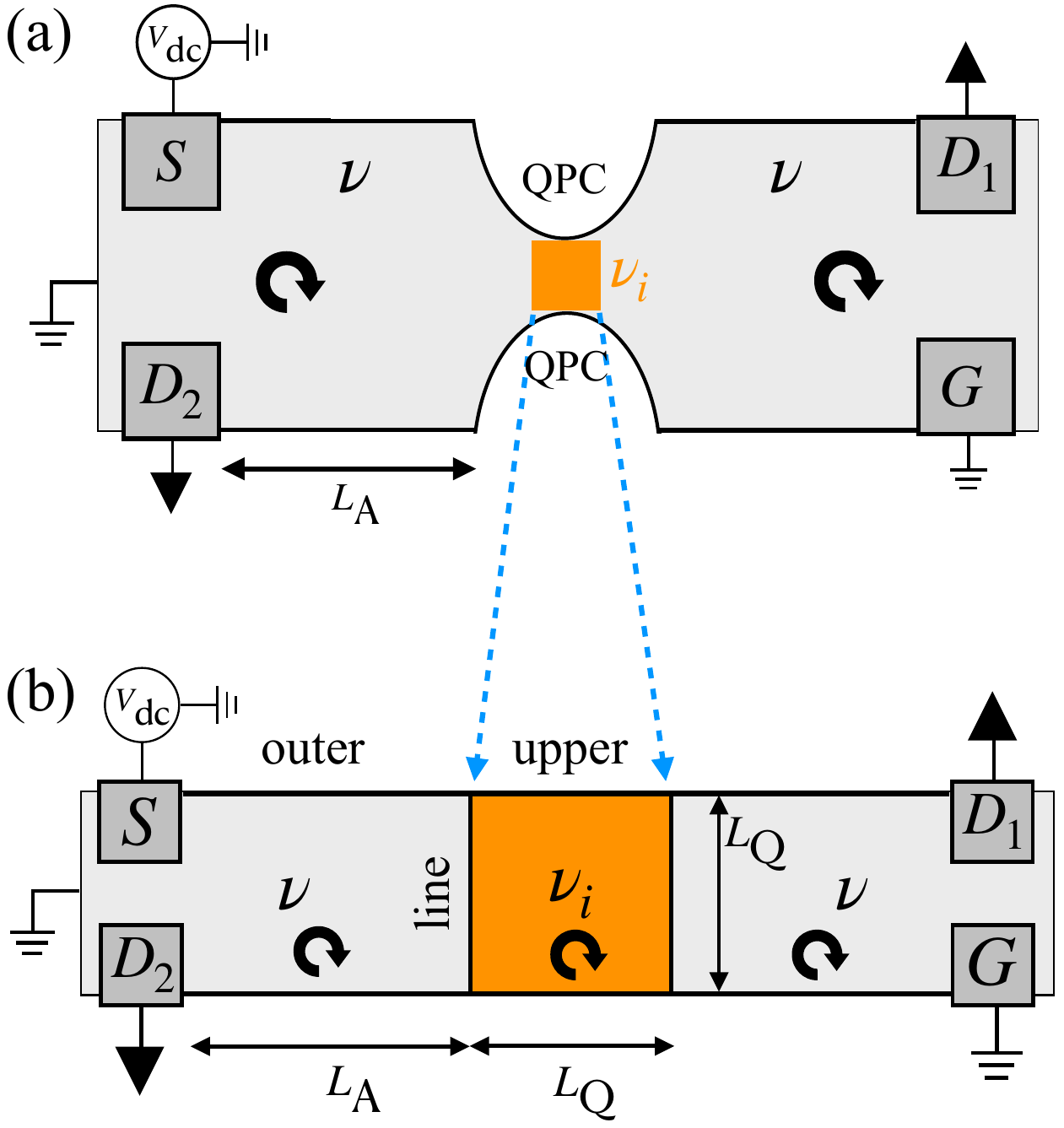}
	\caption{An effective modeling of the QPC
 geometry (c.f.\ \cref{Cartoon}(a)) while we consider equilibration. We describe the chirality of charge propagation of each filling by the circular arrows. (a) We schematically show the QPC geometry, where 
 $\nu$ and $\nu_i (<\nu)$ are the bulk and QPC filling factors respectively and $L_\text{A}, L_\text{Q}$ are the geometric lengths with $L_\text{Q} \ll L_\text{A}$.
 (b) Effective model of
 the QPC geometry with edge equilibration for bulk filling
 $\nu$ and QPC filling $\nu_i$. We show three distinct boundaries: the boundary of $\nu$ with 
 vacuum as ``outer", the boundary of $\nu_i$ with 
 vacuum as ``upper", and the boundary between $\nu$ and
 $\nu_i$ as ``line".} \label{GeometryAndEff}
\end{figure}

Taking into considerations the recent experiments \cite{PhysRevLett.126.216803, Melcer2022, Kumar2022, Srivastav2022}, showing that the thermal equilibration length is order
of magnitude larger than the charge equilibration length, and the typical geometric lengths of a 
QPC geometry (where arm length is much larger than the QPC size) we believe and consider the 
following three distinct relevant possibilities 
(c.f.\ \cref{GeometryAndEff}) :
\begin{enumerate}
\item $l^\text{th}_\text{eq} \ll L_\text{Q}\ll L_\text{A}$ : ``Full" thermal equilibration regime, where the
system is fully thermally equilibrated in every segment.
\item $L_\text{Q} \ll l^\text{th}_\text{eq} \ll L_\text{A}$ : ``Hybrid" thermal equilibration regime, where the
system is fully thermally equilibrated in the arms but thermally unequilibrated around the QPC.
\item $L_\text{Q} \ll L_\text{A} \ll l^\text{th}_\text{eq}$ : ``No" thermal equilibration  regime, where the
system is not thermally equilibrated in any segment.
\end{enumerate}
We will focus our analysis on these aforementioned cases and assume $l^\text{ch}_\text{eq} \ll 
(L_\text{A}, L_\text{Q})$ providing full charge equilibration in each segment of the QPC geometry.

We use Refs.\ \onlinecite{PhysRevB.101.075308, SM5by2, SMshort, SMLong} and write the general expressions for the auto-
and cross-correlations in a QPC geometry for the
bulk filling $\nu$ and QPC filling $\nu_i$ when $\nu>\nu_i$ (\cref{GeometryAndEff} and \cref{ParticleAndHole}).
Charge is assumed to be fully equilibrated, leading to a ballistic charge transport, moving ``downstream" along each segment of the QPC geometry.
We define antiballistic
as the direction opposite to the charge flow (``upstream").
We use those expressions to compute the 
correlation values for 
various choices of $\{\nu,\nu_i\}$ and for different
thermal equilibration regimes under considerations
assuming 
no bulk-leakage
\cite{PhysRevLett.123.137701, Banerjee2018, PhysRevB.99.041302,PhysRevLett.125.157702}.
We write the $\delta^2 I_1, \delta^2 I_2$, and $\delta^2 I_c$ as

\begin{equation}\label{auto1}
	\begin{split}
		\delta^2 I_1 &=2\Bigg(\frac{e^2}{h}\Bigg)\frac{\nu_i}{\nu}(\nu-\nu_i)k_{\text{B}} (T_M+T_N) \\&+ \frac{1}{\nu^2}\Big[\nu_i^2 \langle (\Delta I_S)^2 \rangle + (\nu-\nu_i )^2\langle (\Delta I_G)^2 \rangle\Big],
	\end{split}
\end{equation}
\begin{equation}\label{auto2}
	\begin{split}
		\delta^2 I_2 &=2\Bigg(\frac{e^2}{h}\Bigg)\frac{\nu_i}{\nu}(\nu-\nu_i)k_{\text{B}} (T_M+T_N) \\&+
		\frac{1}{\nu^2}\Big[\nu_i^2 \langle (\Delta I_G)^2 \rangle + (\nu-\nu_i )^2\langle (\Delta I_S)^2 \rangle\Big],
	\end{split}
\end{equation}
\begin{equation}\label{cross}
	\begin{split}
		\delta^2 I_c &=-2\Bigg(\frac{e^2}{h}\Bigg)\frac{\nu_i}{\nu}(\nu-\nu_i)k_{\text{B}} (T_M+T_N) \\&+ \frac{\nu_i(\nu-\nu_i)}{\nu^2}\Big[\langle (\Delta I_G)^2 \rangle + \langle (\Delta I_S)^2 \rangle\Big],
	\end{split}
\end{equation}
where $T_M, T_N$ are the temperatures at the noise spots $M$ and $N$, respectively.
The corresponding Fano factors can be 
found by using \cref{Fano}.
We find $T_M, T_N$ by solving
self-consistent equilibration equations and 
considering energy conservations \cite{SM5by2, SMshort, SMLong, PhysRevB.101.075308}. 
Owing to the Joule heating,
power is dissipated at the hot spots as
\begin{equation}
	\begin{split}
		P_{H_1} = P_{H_2} = \frac{e^2 V_{\text{dc}}^2}{h}\frac{\nu_i(\nu-\nu_i)}{2\nu}.
	\end{split}
\end{equation}

We compute the contributions
$\langle (\Delta I_G)^2 \rangle = \langle (\Delta I_S)^2$
at the noise spots $O$ and $P$ by evaluating the following integral (\cite{SM5by2,PhysRevB.101.075308}) as
\begin{equation}\label{outernoise}
	\begin{split}
	&\langle (\Delta I_S)^2 \rangle = \langle (\Delta I_G)^2 \rangle \\&= \frac{2e^2}{h l_{\text{eq}}^{\text{ch}}}\nu \frac{\nu_{-}}{\nu_{+}}\bigintss_0^{L_{\text{A}}}dx \frac{e^{\frac{-2x}{l_{\text{eq}}^{\text{ch}}}}k_{\text{B}}[T_{+}(x)+T_{-}(x)]}{[1-(e^{\frac{-L_{\text{A}}}{l_{\text{eq}}^{\text{ch}}}}\frac{\nu_{-}}{\nu_{+}})]^2},
	\end{split}
\end{equation}
where $l_{\text{eq}}^{\text{ch}}$ is the charge equilibration length and $\nu=(\nu_{+}-\nu_{-})$. 
We have $\nu_{+}$ and $\nu_{-}$ as the filling factors of the 
downstream and upstream modes, respectively.
Here $T_{\pm}(x)$ are the temperature profiles 
of the modes with filling factors $\nu_{\pm}$, respectively.
We note that $T_{\pm}(x)$ depend on the $l_{\text{eq}}^{\text{th}}$ and we find these by 
following Ref.\ \onlinecite{PhysRevB.101.075308}.
We assume that no voltage drops 
occur along the ``outer" segment (\cref{ParticleAndHole}) and
\begin{equation}
	\begin{split}
		T_{+}(0)=0,\ T_{-}(L_{\text{A}})=T_M
	\end{split}
\end{equation}
serve as the boundary conditions since we consider that the
lead contacts are at zero temperature. 
We point out that, the ``No" thermal equilibration regime was also studied in Ref.\ \onlinecite{Kumar2022} for one edge segment. We note that in this regime, only the upstream modes from
$M$($N$) contribute to $O$($P$), but the downstream modes do not contribute (\cref{ParticleAndHole}(b)). This is
because the downstream modes have zero temperature, which is the contact temperature,
and they are completely thermally
isolated from the upstream modes.
We assume that the particle-hole creation processes (which generate the noise) efficiently transfer heat energy among 
the modes at the noise spot to locally equilibrate those.
This leads to a constant noise in the ``No" thermal equilibration regime (Ref.\ \onlinecite {Park2024} pointed
out a length dependent noise based on a different limit).

\section{Insight into the results}
\label{sec:results}
\SM{
Here, we qualitatively discuss a number of interesting features in the shot noise
Fano factors, which we find by looking at the \cref{tab:CCC_Table} and show their importance.
In the following, we depict those sequentially 
pointing out the usefulness of each of those and 
relating one another:
}

\begin{table}
	\includegraphics[width=\columnwidth]{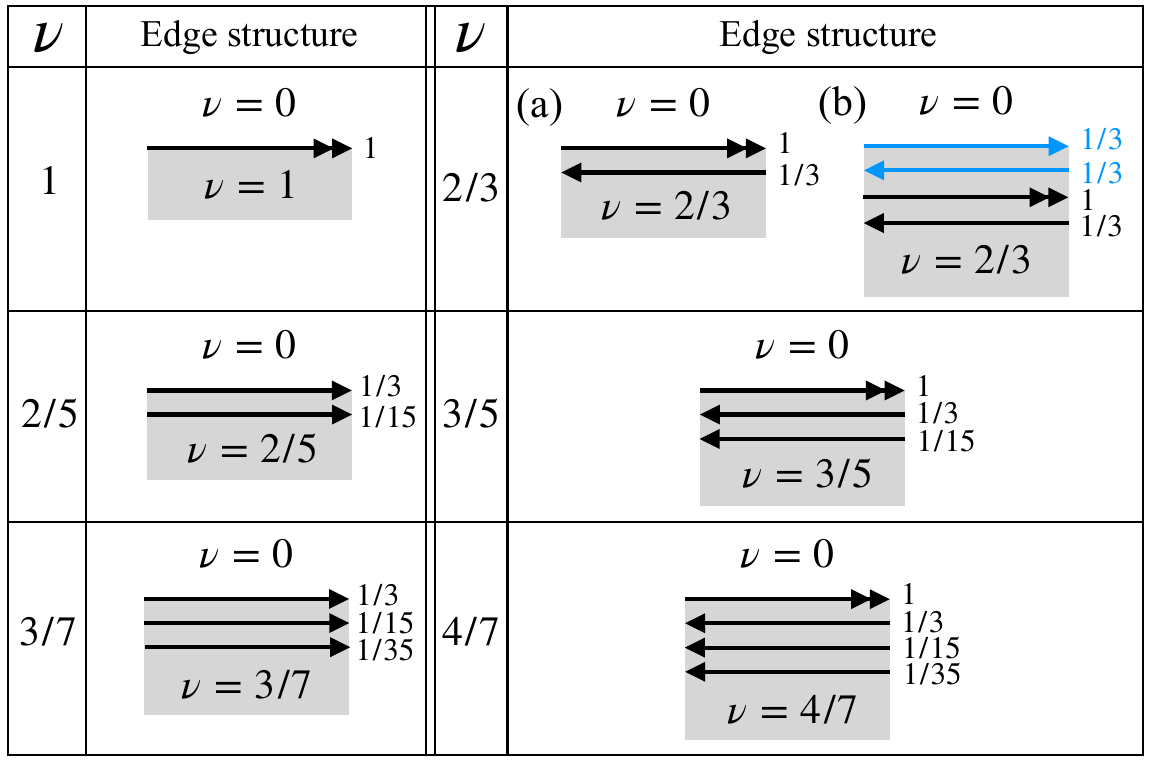}
	\caption{We show the edge structures of integer ($\nu=1$), particle-like ($\nu=1/3, 2/5, 3/7$), and hole-like 
($\nu=2/3, 3/5, 4/7$) filling fractions under 
considerations. Integer and fractional charge modes are depicted as double and single-headed arrows respectively. Each arrow-head shows the chirality of charge propagation of that mode and $\nu=0$ corresponds to the vacuum.
For $\nu=2/3$, we show the bare edge mode structure in (a) and the $1/3$ edge reconstruction (blue) \cite{Wang2013}
in (b).} \label{FFEdges}
\end{table}

\begin{figure}[!t]
	\includegraphics[width=\columnwidth]{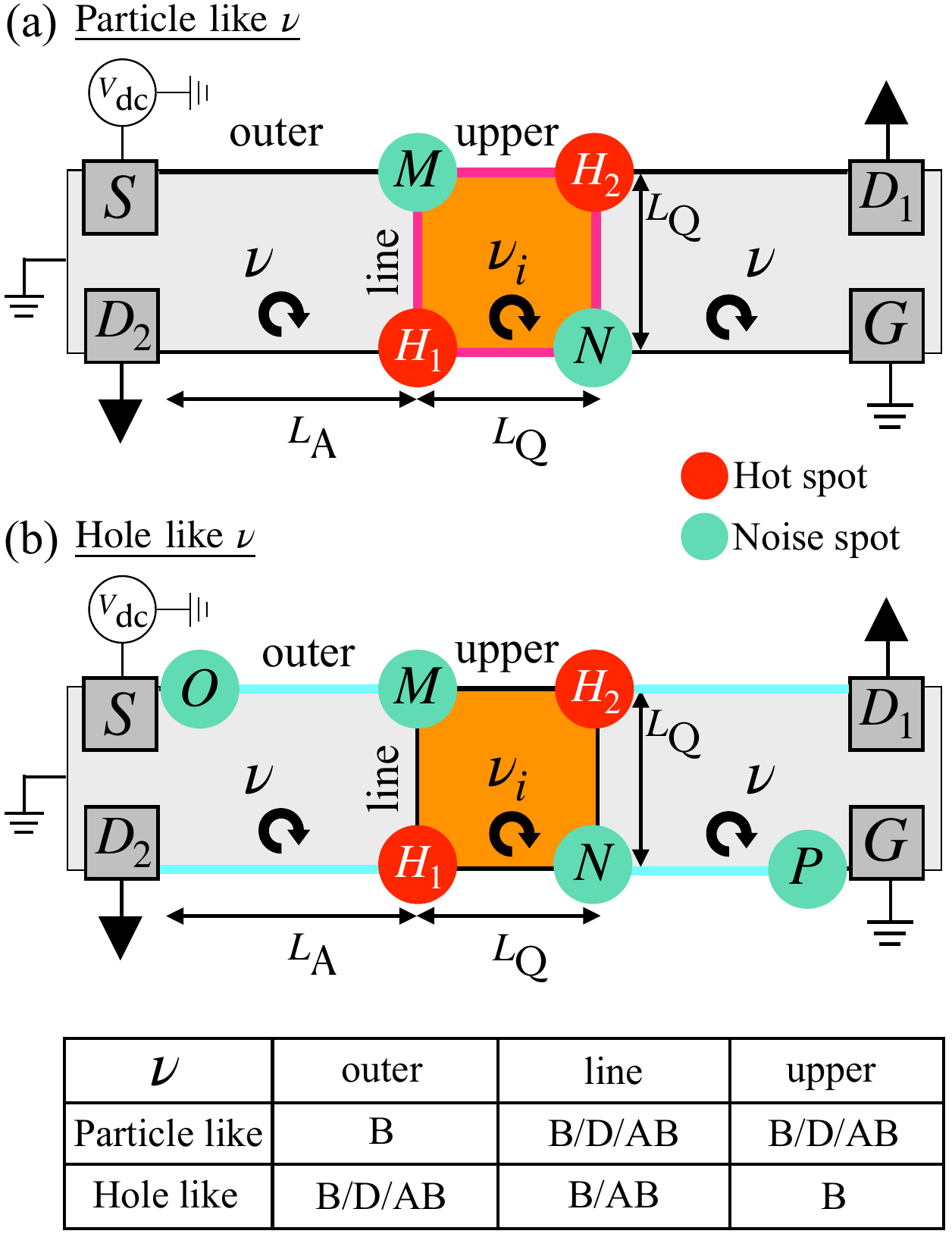}
	\caption{We show the generation of hotspots (red circles)
 resulting in noise spots (green circles) while considering the equilibration
 for both the $\nu=$ particle-like and $\nu=$ hole-like filling fractions
 and the QPC has filling $\nu_i (<\nu)$.
The chirality of charge propagation of each filling is depicted by the circular arrows. (a) We have $\nu$ as the
 particle-like filling fraction. The nature of 
heat transport is ballistic (``B") in the ``outer"
segment and can be ``B" or diffusive (``D")
or anti-ballistic (``AB") in the ``line" and ``upper" segments. We have 
two hot spots as $H_1, H_2$, which result in two noise spots as $M, N$. We note that
the nature of heat transport in the ``line" and ``upper"
segments (pink shaded lines) determine whether we have
a constant noise or length-dependent noise.
(b) We have $\nu$ as the
 hole-like filling fraction. The nature of 
heat transport can be ``B" or ``D" or ``AB" in the ``outer" segment, can be ``B" or ``AB" in the ``line" segment, and is ``B" in the ``upper"
segment. We have two hot spots as $H_1, H_2$,
which result in four noise spots as $M, N, O, P$. 
As there can be ``D" or ``AB" heat transport in the 
``outer" segment, hence we have additional noise spots
$O, P$.
We note that
the nature of heat transport in the ``outer"
segment (cyan shaded lines) determine whether we have
a constant noise or a length-dependent noise.} \label{ParticleAndHole}
\end{figure}

\begin{enumerate}
\item \doubleunderline{$F_1$ vs.\ $F_2$ vs.\ $F_c$} :
\SM{
The present article is based on the study of shot noise, therefore, immediate questions appear to 
a reader's mind regarding the qualitative relations among the
auto- and cross-correlation Fano factors and the 
sign of the cross-correlation.
}
\begin{enumerate}[label=(\Alph*)]
\item \underline{$F_1 = F_2$} :
Two auto-correlations are the same due to the current conservation. 

\item \underline{$F_c < 0$} :
A particle or hole can end up in either $D_1$ or $D_2$, which keeps those drains to be anti-correlated, leading to a negative cross-correlation.
\SM{
Hence, the negative sign of $F_c$ represents that 
the constituents of the noise reach different contacts.
}

\item \underline{$F_1=F_2 \overset{?}{=} -F_c$} :
\SM{
Having understood each of the auto- and cross-correlations individually, a natural question emerges 
as the inequalities among those. In addition, whether
the nature of the bulk filling---being particle-like
or hole-like---can play a role. Below we present a qualitative discussion.
}

We can write \cref{auto1,auto2,cross} in a more simplified manner
as $F_1=F_2=F_M+F_N+F_O+F_P$ and $F_c=-F_M-F_N+F'_O+F'_P$.
If $\langle (\Delta I_S)^2 \rangle = \langle (\Delta I_G)^2 \rangle \approx 0$, then we have 
$F_O=F_P=F'_O=F'_P \approx 0$ leading to $F_1=F_2=-F_c$.
This equality occurs for the $\nu=$ particle-like states 
as they contains co-propagating edges in the 
``outer" segments. Therefore,
heat propagates only ballistically there which leads to the exponentially suppressed contributions from the $O, P$ noise spots (\cref{ParticleAndHole}(a)).

On the contrary, for the $\nu=$ hole-like states
we have counter-propagating edge modes in the 
``outer" segments.
Therefore,
heat can also propagate
diffusively or antiballistically there which lead to constant contributions from the $O, P$ noise spots
as $\langle (\Delta I_S)^2 \rangle = \langle (\Delta I_G)^2 \rangle \neq 0$ (\cref{ParticleAndHole}(b)).
This leads to each of the $F_O, F_P, F'_O, F'_P$ to be
$\neq 0$ and hence $F_1=F_2\neq -F_c$.
\SM{
Such inequalities are indeed found in recent 
experiments \cite{DCG}.
}

\end{enumerate}
\SM{
After learning the inequalities among the Fano factors, naturally we want to understand those
in the quantitative level.
While we give explicit numerical values in \cref{tab:CCC_Table} for the
Fano factors, however, one can qualitatively understand when the values of the Fano factors are
close to zero or a constant or exhibit a length
dependence. We discuss those in the following.
}
\item \doubleunderline{$F_1=F_2=-F_c \approx 0$} :

If we have ballistic heat propagation along each of the 
``outer", ``line", and ``upper" segments, then 
heat can not reach efficiently from any hot spot
to any noise spot. This leads to an 
exponentially suppressed contribution 
to the shot noise from each 
noise spot. Thereby, 
$F_M=F_N=F_O=F_P=F'_O=F'_P \approx 0$ leading to $F_1=F_2=-F_c \approx 0$.

\item  \doubleunderline{$(F_1=F_2,F_c) = \text{constant or\ } f(l^{\text{th}}_{\text{eq}}, L_{\text{Q}}, L_{\text{A}})$ } :

If we have only ballistic or antiballistic
heat transports along each of the segments of the 
QPC geometry, then heat can reach from any hot spot
to any noise spot inefficiently or efficiently, 
respectively. Thereby, a zero or a constant contribution to the shot noise Fano factors
can be attributed.

However, if the heat is transported 
diffusively through any segment then it reaches to 
the contacts very slowly along that segment. This 
gives rise to a length dependence 
in the temperatures of the
hot spots. 
The length dependence has a functional form 
relating the geometric length of that segment
and $l^{\text{th}}_{\text{eq}}$.
The heat reaches 
from the hot spots to the noise spots and thus the noise spots also acquire that length scale dependence in their temperatures, which is manifested in the shot noises.
From \cref{tab:CCC_Table} we note that we have
diffusive heat transports for $\{1,1/3\}$ and 
$\{1,2/3\}$ along ``line" and ``upper" 
segments respectively leading to a 
$\sqrt{l^{\text{th}}_{\text{eq}}/L_{\text{Q}}}$ dependence in the Fano factors.
Similarly, for $\{2/3,1/3\}$ and 
$\{2/3,[1/3]\}$ diffusive heat transports
occur along the ``outer"
segment leading to a 
$\sqrt{L_{\text{A}}/l^{\text{th}}_{\text{eq}}}$ dependence in the Fano factors.

\SM{
The final piece of our observations is to provide
insights on how the bulk filling---being particle-like
or hole-like--- impacts different thermal equilibration regimes which in turn leaves its 
impression on the Fano factors. Below we show that 
interesting features arise and one can understand 
those in a qualitative level, keeping numerical values aside.
}

\item \doubleunderline{``No" = ``Hybrid" $\neq$ ``Full" for $\nu=$ particle-like} :

The difference between ``No" and ``Hybrid"
thermal equilibration regimes is that 
all the edge modes remain thermally unequilibrated in the
former while they equilibrate in the latter only along
the ``outer" segment (c.f.\ \cref{ParticleAndHole}(a)). Now, for the $\nu=$ particle-like states the heat transport along the ``outer" segment is 
ballistic always since we have co-propagating modes.
Hence, it does not matter if those modes in the 
``outer" segment thermally equilibrate or not, since they 
will not contribute to the shot noise leading to 
``No" = ``Hybrid".
However, in the ``Full" thermal equilibration regime the modes 
thermally equilibrate both along the 
``line" and ``upper" segments (c.f.\ \cref{ParticleAndHole}(a)) leading to 
``No" = ``Hybrid" $\neq$ ``Full".

\item \doubleunderline{``No" $\neq$ ``Hybrid" = ``Full" for $\nu=$ hole-like} :

A common fact between the ``Hybrid" and ``Full"
thermal equilibration regimes is that
the ``outer" segment is thermally 
equilibrated for both and
the heat transport along the ``outer" segment is 
antiballistic (c.f.\ \cref{ParticleAndHole}(b)). Therefore, no heat can reach to the
drains from a hot spot.
It does not matter if the ``line" and ``upper"
segments (c.f.\ \cref{ParticleAndHole}(b)) are thermally equilibrated (``Full") or not (``Hybrid"),
all the heat is transferred from a hot spot to the noise spots leading to ``Hybrid" = ``Full".
However, in the ``No" thermal equilibration regime the modes are not thermally equilibrated
along the ``outer" segment. Hence the ballistic modes, connecting a hot spot to its nearest
drain, leak  some amount of heat to that drain leading to ``No" $\neq$ ``Hybrid" = ``Full".
\end{enumerate}

\section{Importance of a ``Hybrid" thermal equilibration regime}
The ``Hybrid" thermal equilibration regime—defined by the separation of length scales
$L_\text{Q} \ll l^\text{th}_\text{eq} \ll L_\text{A}$—is a generic operating point of present-day QPC devices once we accept two robust experimental facts:
(i) charge equilibrates much faster than heat ($l^\text{ch}_\text{eq} \ll l^\text{th}_\text{eq}$)
and (ii) device arms are much longer than the QPC constriction ($L_\text{Q} \ll L_\text{A}$).
In this window, charge is fully equilibrated everywhere, while heat equilibrates in the arms (``outer" segments) but not in the short region surrounding the QPC (``line"/``upper" segments).
This spatially inhomogeneous equilibration is precisely what the ``Hybrid" thermal
equilibration regime encodes.
It equips us with a controlled way to separate local QPC physics from long-arm heat transport, turning shot-noise at a conductance plateau into a nonlocal probe of edge hydrodynamics.

In particular, this raises a major concern in the experimental platform, asking the question of how deeply this regime affects the different measurement outcomes. We here provide a concrete answer and a principle that dictates that answer for the two different types of edges, namely particle-like edges and hole-like edges. In particle-like edges (co-propagating ``outer" modes), ``Hybrid" and ``No" thermal equilibration give the same shot-noise outcomes, revealing that arm thermal equilibration is inert while QPC-region thermal equilibration matters. Conversely, in hole-like edges (counter-propagating ``outer" modes), ``Hybrid" and ``Full" thermal equilibration coincide, exposing the arms as the relevant thermal bottleneck.
Thus, the `Hybrid" thermal equilibration regime sorts whether noise is controlled by the arms or by the QPC segment, which is also relevant for experiments involving
QPC devices.
Moreover, the ``Hybrid" thermal
equilibration regime
asks future theory and experiment to treat thermal equilibration as segment-selective rather than uniform.

\section{Comparison with recent experiments}
\SM{
Our study assumes full charge equilibration in each
segment of the QPC geometry, giving rise to well
quantized QPC conductance plateaus. 
Such formation of QPC plateaus are robust and quite generally backed up by recent experiments \cite{Bid2009,Bhattacharyya2019,Sabo2017,Zimmermann2017,Pandey2024,Manfra2022,Hirayama2023,roy2025halfintegerthermalconductanceabsence}.
In this paper, we are only dealing with $\nu_i<\nu$ leading to a QPC plateau at $\nu_i/\nu$, however, 
cases with $\nu_i>\nu$ can also lead to a QPC plateau
following charge equilibration as shown in Refs.\ 
\onlinecite{SMLong,SMshort}. We stress that such 
mechanisms leading to quantized plateaus are not only
tied with 2DEG systems \cite{Bid2009,Bhattacharyya2019,Sabo2017,Manfra2022,Hirayama2023} but also with recent studies in graphene
systems \cite{Zimmermann2017,Pandey2024}.

These plateaus are in general noisy, which is qualitatively consistent with Ref.\ \onlinecite{Bhattacharyya2019}. However, the auto-correlation Fano factors were found to be the same as the bulk filling
in Ref.\ \onlinecite{Bhattacharyya2019}. In our 
study, for some cases, we have found the Fano factors
are comparable but somewhat larger than in Ref.\ \onlinecite{Bhattacharyya2019}. For example, in the
case $(\nu,\nu_i)={3/5,1/3}$, we have $F_1=F_2\approx 0.53$ in ``No" and ``Hybrid" thermal equilibration regimes and in the
case $(\nu,\nu_i)={4/7,1/3}$, we have $F_1=F_2\approx 0.39$ in ``No", ``Hybrid", and ``Full" thermal equilibration regimes (\cref{tab:CCC_Table}). The measurement of cross-
correlations can distinguish different thermal equilibration regimes, while the auto-correlations are same (\cref{tab:CCC_Table}). We note a feature in 
our \cref{tab:CCC_Table} for hole-like states and
we point out that this is indeed consistent with 
recent experiments \cite{DCG}.}

\section{Summary and future prospects}
\label{sec:summary}
In this work, we have shown how the latest findings \cite{PhysRevLett.126.216803, Melcer2022, Kumar2022, Srivastav2022} of the
inequality between charge and thermal equilibration lengths can affect the electrical shot noise. 
Our findings depict that distinct universal facets arise among the different
thermal equilibration regimes for the edges of particle-like and
and hole-like fillings fractions.
We have analyzed different
Abelian quantum Hall states as the bulk filling ($\nu$) and their known or possible QPC transmission plateaus (\cref{FFEdges}). We have assumed full charge equilibration throughout our analysis, which has licensed us
in determining the transmission plateau
for a given QPC filling ($\nu_i<\nu$). We have considered three
different inequalities among the
geometric lengths ($L_\text{A}, L_\text{Q}$) and the thermal equilibration
lengths ($l^\text{th}_\text{eq}$).
These three regimes of thermal equilibration 
have been denoted as 
``No" (both the arm and the edges around QPC are thermally unequilibrated),
``Hybrid" (the arm is thermally equilibrated but not the edges around QPC), and
``Full" (both the arm and the edges around QPC are thermally equilibrated).
We have found for the particle-like states the thermal equilibration of the 
edges around the QPC affects the shot noise and for the hole-like states the
thermal equilibration of the edges at the arms affects the shot noise (c.f.\ \cref{tab:CCC_Table} and \cref{ParticleAndHole}). We have found that 
the magnitude of the auto-correlation Fano factors is always larger than the cross-correlation Fano factor for the hole-like states, since the thermal
transport 
contribution along the arm is either diffusive
or antiballistic.

This work is a step in understanding 
how the shot noise is affected by different
thermal equilibration regimes in a QPC geometry
at a QPC transmission plateau. We note
that the analyses are valid if we consider 
similar geometry made out of interfaces \cite{SM5by2}.
This work can be extended beyond the Abelian 
quantum Hall states
to the non-Abelian fillings (e.g. $\nu=12/5$ \cite{ReadRezayi1999,yutushui2023}), where there are no extensive studies yet (recently
$\nu=5/2$ is considered \cite{SM5by2}).
In this work, we only focus on understanding the conventional QPC transmission plateau without
the engineered interface modes. 
However, 2DEG interface QPC \cite{biswas2023} is an intriguing direction that we hope to look into in the future.
Another interesting direction will be the graphene quantum Hall effect. With the recent
improvements and understandings, QPC and interfaces in graphene quantum Hall systems have become
very relevant \cite{Ronen_2021,paul2022}. 
We also note that in graphene how to construct
a QPC is not unique, which makes the problem even richer.
Interesting future directions may include other
topological insulators (for example quantum spin Hall states) \cite{PhysRevLett.128.186801, PhysRevLett.118.046801, John_2022}, quantum spin liquids \cite{Kitaev2006, Banerjee2016, Kasahara2018}, and more.

\begin{acknowledgements}
We thank Moshe Goldstein and Yuval Gefen for many illuminating
discussions. We thank Udit
Khanna for useful discussions.
S.M.\ was supported by the Weizmann Institute of Science, Israel Deans fellowship through Feinberg
Graduate School and the senior postdoctoral fellowship, as well as the Raymond Beverly Sackler Center for Computational Molecular and Material Science at Tel Aviv University. 
A.D.\ was supported by DFG
MI 658/10-2 and DFG RO 2247/11-1. A.D.\ also thanks the Israel Planning and budgeting committee (PBC) and the
Weizmann Institute of Science, the Dean of Faculty fellowship, and the Koshland Foundation for financial support. A.D.\ thanks IISER Tirupati start up grant for support.
\end{acknowledgements}

\appendix
\SM{\section{Derivations of $\delta^2 I_1, \delta^2 I_2, \delta^2 I_c$}}

\SM{We follow Refs.\ \onlinecite{PhysRevB.101.075308, SMshort,SMLong,SM5by2} and re-derive the expressions of
$\delta^2 I_1, \delta^2 I_2, \delta^2 I_c$ explaining each term
in full details which will
help to follow our entire calculations in the main manuscript. We assume that
the geometric lengths are larger
than the charge equilibration length leading to
$(L_{\text{A}}, L_{\text{Q}})\gg l^\text{ch}_\text{eq}$.}

\begin{figure}[H]
	\includegraphics[width=0.95\columnwidth]{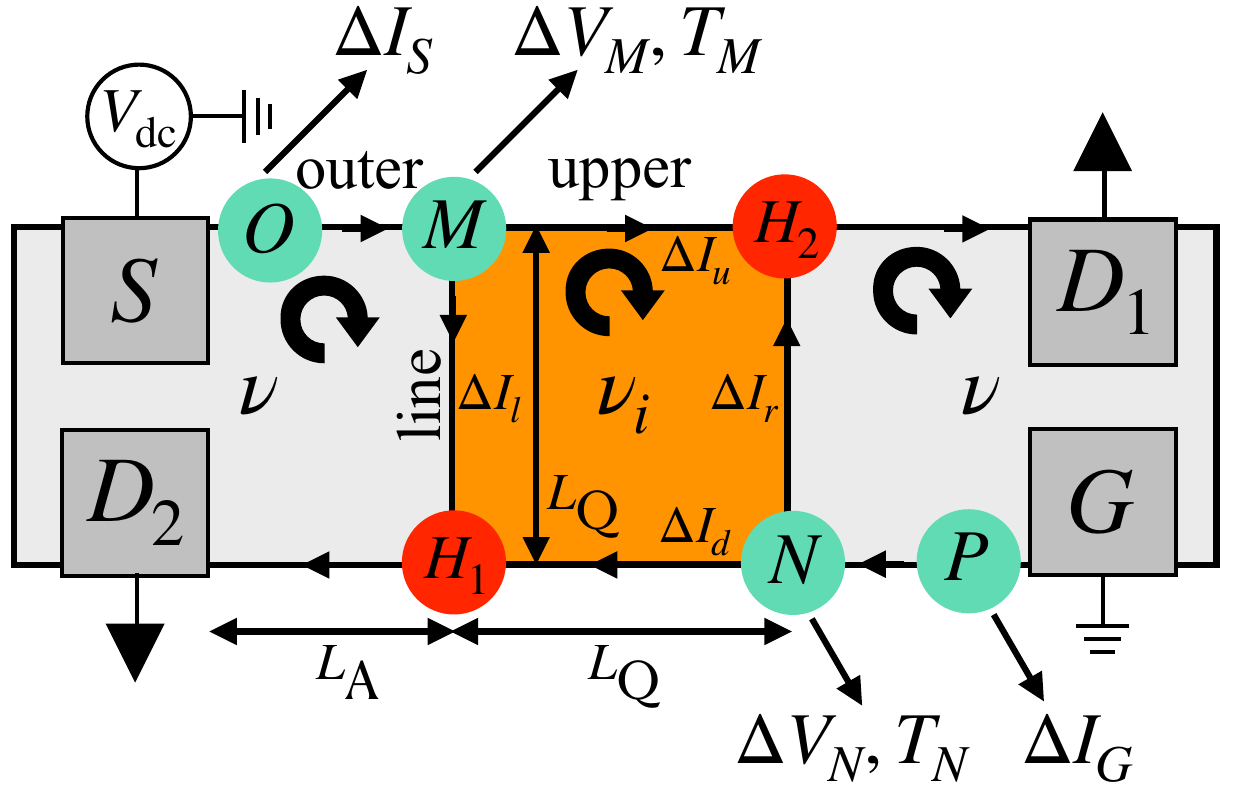}
	\caption{\SM{A Hall bar device with bulk filling $\nu$ 
    and QPC filling $\nu_i(< \nu)$ with 
 a source $S$ (biased by a dc voltage $V_{\text{dc}}$), a ground contact $G$, and two drains, $D_1$ and $D_2$. The geometric lengths are $L_{\text{A}}, L_{\text{Q}}$. We denote the segment between the vacuum and $\nu$ as  ``outer", between the vacuum and $\nu_i$ as  ``upper", and between $\nu$ and $\nu_i$ as ``line". The arrow
 shows the fully
 equilibrated charge propagation along each segment while the circular arrow depicts the chirality of each filling.
 Hot spots $H_1,H_2$ (red circles) are created due to the
 voltage drops which are responsible for the creation of noise spots $M,N,O,P$ (green circles) \cite{PhysRevB.101.075308}.}}\label{FigAppendix}
\end{figure} 

\begin{figure}[h!]
	\includegraphics[width=\columnwidth]{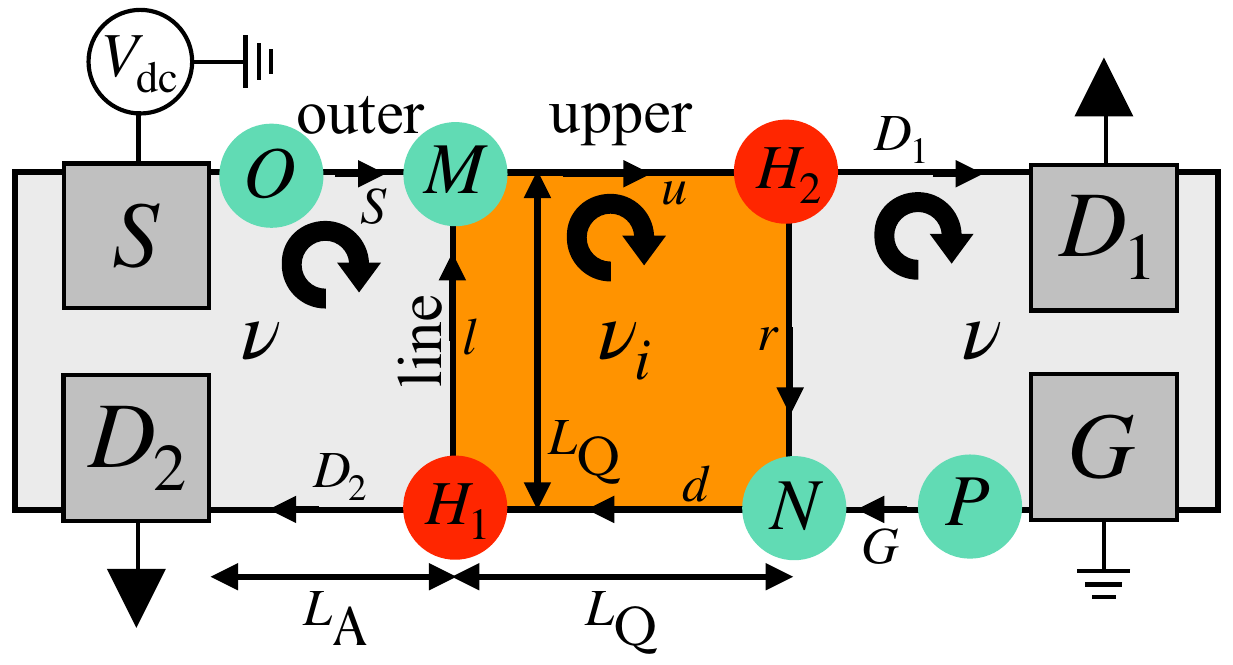}
	\caption{\SM{We assume charge to be fully equilibrated leading to ballistic transport, moving ``downstream" along each segment of the device. Each circular arrow shows the chirality of charge
propagation of the respective filling ($\nu>\nu_i$). The direction opposite to charge flow (``upstream") is taken as antiballistic. Heat is also fully 
equilibrated and we show here the case when the heat transport is ballistic in ``outer", antiballistic in ``line", and ballistic in
``upper" segments, as shown by the arrows. Voltage drops occur at the hot spots $H_1,H_2$ (red circles), giving rise to the noise spots $M,N,O,P$ (green circles) \cite{PhysRevB.101.075308}.}
} \label{heat}
\end{figure}

\SM{We write the following equations for the current fluctuations as
\begin{equation}
	\begin{split}
		\Delta I_1 &= \Delta I_u + \Delta I_r,\\ \Delta I_2 &= \Delta I_d + \Delta I_l,\\
  \Delta I_S &= \Delta I_u + \Delta I_l,\\
  \Delta I_G &= \Delta I_d + \Delta I_r,
	\end{split}
\end{equation}
and
\begin{equation}
	\begin{split}
		& \Delta I_l = (\nu-\nu_i) \frac{e^2}{h} \Delta V_M + \Delta I_l^{\text{th}},\\& \Delta I_u = \nu_i \frac{e^2}{h} \Delta V_M + \Delta I_u^{\text{th}},\\&\Delta I_r = (\nu-\nu_i) \frac{e^2}{h} \Delta V_N + \Delta I_r^{\text{th}},\\& \Delta I_d = \nu_i \frac{e^2}{h} \Delta V_N + \Delta I_r^{\text{th}}.
	\end{split}
\end{equation}
Hence, we find
\begin{equation}
	\begin{split}
		\Delta I_1 &= \nu_i \frac{e^2}{h} \Delta V_M + \Delta I_u^{\text{th}}+(\nu-\nu_i) \frac{e^2}{h} \Delta V_N + \Delta I_r^{\text{th}}\\&=\frac{\nu_i}{\nu}\Delta I_S +
		\frac{(\nu-\nu_i)}{\nu}\Delta I_G \\&+ \frac{(\nu-\nu_i)}{\nu}(\Delta I_u^{\text{th}} - \Delta I_d^{\text{th}})+\frac{\nu_i}{\nu}(\Delta I_r^{\text{th}} - \Delta I_l^{\text{th}})
	\end{split}
\end{equation}
and
\begin{equation}
	\begin{split}
		\Delta I_2 &=\frac{\nu_i}{\nu}\Delta I_G + \frac{(\nu-\nu_i)}{\nu}\Delta I_S \\&+
		\frac{(\nu-\nu_i)}{\nu}(\Delta I_d^{\text{th}} - \Delta I_u^{\text{th}})+\frac{\nu_i}{\nu}(\Delta I_l^{\text{th}} - \Delta I_r^{\text{th}}).
	\end{split}
\end{equation}
We use the local Johnson-Nyquist relations for thermal noise as
\begin{equation}
	\begin{split}
		&\ \ \ \ \  \langle (\Delta I_l^{\text{th}})^2 \rangle = \frac{2 e^2}{h} (\nu-\nu_i) k_{\text{B}} T_M,\\& \ \ \ \ \ \langle (\Delta I_u^{\text{th}})^2 \rangle = \frac{2 e^2}{h} \nu_i k_{\text{B}} T_M,\\& \ \ \ \ \ \langle (\Delta I_r^{\text{th}})^2 \rangle = \frac{2 e^2}{h}
		(\nu-\nu_i)k_{\text{B}} T_N,\\& \ \ \ \ \ \langle (\Delta I_d^{\text{th}})^2 \rangle = \frac{2 e^2}{h} \nu_i k_{\text{B}} T_N,\\& \langle (\Delta I_i^{\text{th}} \Delta I_j^{\text{th}}) \rangle = 0,\ \text{for}\ i \neq j\ \text{and}\ i,j \in \{l,u,r,d\},
	\end{split}
\end{equation}
and find the correlations to be
\begin{equation}
	\begin{split}
		\delta^2 I_1 &=2\Bigg(\frac{e^2}{h}\Bigg)\frac{\nu_i}{\nu}(\nu-\nu_i)k_{\text{B}} (T_M+T_N) \\&+ \frac{1}{\nu^2}\Big[\nu_i^2 \langle (\Delta I_S)^2 \rangle + (\nu-\nu_i )^2\langle (\Delta I_G)^2 \rangle\Big],
	\end{split}
\end{equation}
\begin{equation}
	\begin{split}
		\delta^2 I_2 &=2\Bigg(\frac{e^2}{h}\Bigg)\frac{\nu_i}{\nu}(\nu-\nu_i)k_{\text{B}} (T_M+T_N) \\&+
		\frac{1}{\nu^2}\Big[\nu_i^2 \langle (\Delta I_G)^2 \rangle + (\nu-\nu_i )^2\langle (\Delta I_S)^2 \rangle\Big],
	\end{split}
\end{equation}
and
\begin{equation}
	\begin{split}
		\delta^2 I_c &=-2\Bigg(\frac{e^2}{h}\Bigg)\frac{\nu_i}{\nu}(\nu-\nu_i)k_{\text{B}} (T_M+T_N) \\&+ \frac{\nu_i(\nu-\nu_i)}{\nu^2}\Big[\langle (\Delta I_G)^2 \rangle + \langle (\Delta I_S)^2 \rangle\Big],
	\end{split}
\end{equation}
where $T_M, T_N$ are, respectively, the temperatures at the noise spots $M$ and $N$. The computation of $T_M, T_N$ requires the
nature of heat transport along each segment of the device. Below we 
consider a particular case to show how $T_M, T_N$ are calculated 
by solving
self-consistent equilibration equations and 
considering energy conservation at each edge mode junction \cite{SM5by2,PhysRevB.101.075308}.

\section{Computation of $T_M, T_N$}

We show the detailed calculation of $T_M, T_N$ when charge 
and heat are fully equilibrated in each segment of the device (leading to
$(L_{\text{A}}, L_{\text{Q}})\gg l^\text{ch}_\text{eq}$
and $(L_{\text{A}}, L_{\text{Q}})\gg l^\text{th}_\text{eq}$) and the heat transport is ballistic in ``outer", antiballistic in ``line", and ballistic in
``upper" segments (\cref{heat}). Other cases can be derived in a similar manner.}

\SM{We calculate $T_M, T_N$ assuming zero temperature in the contacts/leads. We write the energy conservation equations at
$H_1, N, H_2, M$ as \cite{PhysRevB.101.075308}
\begin{equation}\label{heatcurrent}
	\begin{split}
		&J_{D_2} + J_l = P_{H_1} + J_d,\\&J_G+J_r=J_d,\\&J_{D_1}+J_r=P_{H_2}+J_u,\\&J_S+J_l=J_u,
	\end{split}
\end{equation}
where we have $J_i$ as the heat current along $i \in \{ S, G, D_1, D_2, u,r,d,l \}$th segment, and $P_{H_1}, P_{H_2}$ as, respectively, the powers dissipated in the hot spots $H_1, H_2$.

We write the following expressions as
\begin{equation}\label{heatcurrentBABB}
	\begin{split}
		&J_S=0,\ J_u=\frac{\kappa}{2}T_M^2(\delta c)_u,\ J_{D_1}=\frac{\kappa}{2}T_{H_2}^2(\delta c)_1,\\&J_G=0,\ J_d=\frac{\kappa}{2}T_N^2(\delta c)_d,\ J_{D_2}=\frac{\kappa}{2}T_{H_1}^2(\delta c)_2,\\&J_l=\frac{\kappa}{2}T_{H_1}^2(\delta c)_l,\ J_r=\frac{\kappa}{2}T_{H_2}^2(\delta c)_r,
	\end{split}
\end{equation}
where 
$\kappa = \frac{\pi^2 k_{\text{B}}^2}{3h}$. Here, $T_M, T_N, T_{H_1}, T_{H_2}$ are the temperatures at
$M, N, H_1, H_2$ respectively.
$P_{H_1}, P_{H_2}$ are the powers dissipated due to the Joule heating corresponding to the associated voltage drop at $H_1, H_2$, respectively. We derive its explicit expression later
and find $P_{H_1}=P_{H_2}$. We have  $(\delta c)_i = |(c_{\text{down}}-c_{\text{up}})_i|$ as the modulus of difference between the central charges of downstream and upstream modes in the $i$th segment.
We write 
\begin{equation}
	\begin{split}
		&\frac{\kappa}{2}T_{H_1}^2(\delta c)_2+\frac{\kappa}{2}T_{H_1}^2(\delta c)_l=P_{H_1}+\frac{\kappa}{2}T_N^2(\delta c)_d,\\&0+\frac{\kappa}{2}T_{H_2}^2(\delta c)_r=\frac{\kappa}{2}T_N^2(\delta c)_d,\\&\frac{\kappa}
        {2}T_{H_2}^2(\delta c)_1+\frac{\kappa}{2}T_{H_2}^2(\delta c)_r=P_{H_2}+\frac{\kappa}{2}T_M^2(\delta c)_u,\\&0+\frac{\kappa}{2}T_{H_1}^2(\delta c)_l=\frac{\kappa}{2}T_M^2(\delta c)_u,
	\end{split}
\end{equation}
and therefore,
\begin{equation}
	\begin{split}
		&\frac{[(\delta c)_1 + (\delta c)_r]}{(\delta c)_r}(\delta c)_d\frac{\kappa}{2}T_N^2=P_{H_2}+\frac{\kappa}{2}(\delta c)_uT_M^2,\\&\frac{[(\delta c)_2 + (\delta c)_l]}{(\delta c)_l}(\delta c)_u\frac{\kappa}{2}T_M^2=P_{H_1}+\frac{\kappa}{2}(\delta c)_dT_N^2.
	\end{split}
\end{equation}
This leads to 
\begin{equation}
	\begin{split}
		&\frac{\kappa}{2}(\delta c)_uT_M^2\Bigg[ 1+\frac{(\delta c)_2+(\delta c)_l}{(\delta c)_l} \Bigg] \\&= \frac{\kappa}{2}(\delta c)_dT_N^2\Bigg[ 1+\frac{(\delta c)_1+(\delta c)_r}{(\delta c)_r} \Bigg].
	\end{split}
\end{equation}
We have $(\delta c)_u=(\delta c)_d,\ (\delta c)_2=(\delta c)_1,\ (\delta c)_l=(\delta c)_r$, and hence $T_M=T_N$. 
Thereby,
\begin{equation}
	\begin{split}
		\frac{\kappa T_M^2}{2}\frac{(\delta c)_1}{(\delta c)_r}(\delta c)_d=P_{H_2},
	\end{split}
\end{equation}
leading to
\begin{equation}\label{TcBABB}
	\begin{split}
		T_M=T_N=\sqrt{\frac{2 P_{H_2}}{\kappa}\frac{(\delta c)_r}{(\delta c)_1 (\delta c)_d}}.
	\end{split}
\end{equation}

\section{Computation of $P_{H_1}=P_{H_2}$}

We refer to \cref{FigAppendix} (assuming that the geometric lengths are larger
than the charge equilibration length leading to
$(L_{\text{A}}, L_{\text{Q}})\gg l^\text{ch}_\text{eq}$) and write the electrical
current conservation equations at $M,N,H_1,H_2$ respectively as

\begin{equation}
\begin{split}
&V_{\text{dc}}=V_M,\\&V _N=0,\\&(\nu-\nu_i)V_M=\nu V_{H_1},\\&
\nu_i V_M=\nu V_{H_2}.
\end{split}
\end{equation}

We find 

\begin{equation}
\begin{split}
V_{H_1}=\frac{(\nu-\nu_i)}{\nu}V_{\text{dc}}, V_{H_2}=\frac{\nu_i}{\nu}V_{\text{dc}}.
\end{split}
\end{equation}

The dissipated power at $H_1$ is

\begin{equation}
\begin{split}
P_{H_1}&=\frac{e^2}{2h}\Bigg[ (\nu-\nu_i) V_M^2 + \nu_i^2 V_N^2-\nu V_{H_1}^2 ]\\&=\frac{e^2 V_{\text{dc}}^2}{h}\frac{\nu_i(\nu-\nu_i)}{2\nu}.
\end{split}
\end{equation}

Similar calculations show $P_{H_1}=P_{H_2}$.

\section{Calculation of $\langle (\Delta I_S)^2 \rangle=\langle (\Delta I_G)^2 \rangle$}

We consider two counter-propagating modes in a quantum Hall edge, where inter-mode charge tunneling happens. Shot noise occurs while electric current is
partitioned. The fluctuation of local tunneling current has two parts as

\begin{equation}
    \Delta I^\tau_j=\Delta I^{\tau,th}_j+\Delta I^{\tau,v}_j.
\end{equation}

Here, $\Delta I^{\tau,th}_j$ is the fluctuation for the local Johnson-Nyquist noise with 

\begin{equation}
    \langle \Delta I^{\tau,th}_j \Delta I^{\tau,th}_j \rangle
    =\frac{2 e^2}{h}gk_{\text{B}}(T_{+,j}+T_{-,j}),
\end{equation}
where $g$ is the tunneling coupling between the two modes having
local temperatures $T_{+,j}$ and $T_{-,j}$ at $j$th location.

We have
\begin{equation}
    \Delta I^{\tau,v}_j=g\frac{e^2}{h}(\Delta V_{+,j}-\Delta V_{-,j}),
\end{equation}
where the two modes having
local voltages $V_{+,j}$ and $V_{-,j}$ at $j$th location.

We take the continuum limit and find the expression for the noise 
profile \cite{PhysRevB.99.161302, PhysRevB.101.075308} as written in \cref{outernoise} of the main text.

\section{Temperature profiles $T_{+}(x), T_{-}(x)$}

The local temperature profiles of the modes are computed by solving the
following differential equation \cite{PhysRevB.101.075308}.

\begin{equation}
    \partial_x \begin{pmatrix}
T^2_+(x) \\
T^2_-(x) 
\end{pmatrix} \sim \frac{1}{l} \begin{pmatrix}
-n_u&n_u \\
-n_d&n_d 
\end{pmatrix}\begin{pmatrix}
T^2_+(x) \\
T^2_-(x) 
\end{pmatrix}+P_J,
\end{equation}
where $P_J$ is the power dissipation due to Joule heating and 
$n_u, n_d$ are respectively the total number of upstream and downstream modes
and $l \propto l^\text{th}_\text{eq}$. We have distinct solutions of this equation for different heat
transport---ballistic, antiballistic, diffusive. These distinct solutions, while inserted in \cref{heat}, lead to different noise profiles.}

\bibliography{references}
\end{document}